\documentclass{article}

\usepackage{microtype}
\usepackage{graphicx}
\usepackage{subfigure}
\usepackage{booktabs} 
\usepackage[small,compact]{titlesec}

\usepackage{hyperref}

\usepackage[accepted]{sysml2019}

\usepackage{array}
\newcolumntype{P}[1]{>{\centering\arraybackslash}p{#1}}
\newcolumntype{M}[1]{>{\centering\arraybackslash}m{#1}}

\sysmltitlerunning{Benchmarking TinyML Systems}

\begin{document}

\twocolumn[
\sysmltitle{Benchmarking TinyML Systems: Challenges and Direction }



\sysmlsetsymbol{equal}{*}

\begin{sysmlauthorlist}
\sysmlauthor{Colby R. Banbury}{hvd}
\sysmlauthor{Vijay Janapa Reddi}{hvd}
\sysmlauthor{Max Lam}{hvd}
\sysmlauthor{William Fu}{hvd}
\sysmlauthor{Amin Fazel}{Samsung}
\sysmlauthor{Jeremy Holleman}{syn,uncc}
\sysmlauthor{Xinyuan Huang}{cis}
\sysmlauthor{Robert Hurtado}{California State Polytechnic University}
\sysmlauthor{David Kanter}{Real World Insights}
\sysmlauthor{Anton Lokhmotov}{dividiti}
\sysmlauthor{David Patterson}{ucb,google}
\sysmlauthor{Danilo Pau}{STM}
\sysmlauthor{Jae-sun Seo}{Arizona State University}
\sysmlauthor{Jeff Sieracki}{Reality AI}
\sysmlauthor{Urmish Thakker}{Arm ML Research Lab}
\sysmlauthor{Marian Verhelst}{kul,imec}
\sysmlauthor{Poonam Yadav}{University of York}

\end{sysmlauthorlist}

\sysmlaffiliation{hvd}{Harvard University}
\sysmlaffiliation{syn}{Syntiant}
\sysmlaffiliation{uncc}{University of North Carolina, Charlotte}
\sysmlaffiliation{cis}{Cisco Systems}
\sysmlaffiliation{kul}{KU Leuven}
\sysmlaffiliation{imec}{Interuniversity Microelectronics Centre (IMEC)}
\sysmlaffiliation{STM}{STMicroelectronics, Italy}
\sysmlaffiliation{dividiti}{dividiti}
\sysmlaffiliation{Samsung}{Samsung Semiconductor, Inc.}
\sysmlaffiliation{Reality AI}{Reality AI}
\sysmlaffiliation{Arizona State University}{Arizona State University}
\sysmlaffiliation{University of York}{University of York}
\sysmlaffiliation{California State Polytechnic University}{California State Polytechnic University, Pomona}
\sysmlaffiliation{Arm ML Research Lab}{Arm ML Research Lab}
\sysmlaffiliation{Real World Insights}{Real World Insights}
\sysmlaffiliation{ucb}{University of California, Berkeley}
\sysmlaffiliation{google}{Google}

\sysmlcorrespondingauthor{Colby R. Banbury}{cbanbury@g.harvard.edu}

\sysmlkeywords{Machine Learning, SysML}

\vskip 0.3in

\begin{abstract}
Recent advancements in ultra-low-power machine learning (TinyML) hardware promises to unlock an entirely new class of smart applications. However, continued progress is limited by the lack of a widely accepted benchmark for these systems. Benchmarking allows us to measure and thereby systematically compare, evaluate, and improve the performance of systems and is therefore fundamental to a field reaching maturity. In this position paper, we present the current landscape of TinyML and discuss the challenges and direction towards developing a fair and useful hardware benchmark for TinyML workloads. Furthermore, we present our four benchmarks and discuss our selection methodology.
Our viewpoints reflect the collective thoughts of the TinyMLPerf working group that is comprised of over 30 organizations.
\end{abstract}]


\printAffiliationsAndNotice{}  

\section{Introduction}
Machine learning (ML) inference on the edge is an increasingly attractive prospect due to its potential for increasing energy efficiency~\cite{sparse}, privacy, responsiveness~\cite{hello-edge}, and autonomy of edge devices.
Thus far, the field edge ML has predominately focused on mobile inference which has led to numerous advancements in machine learning models such as exploiting pruning, sparsity, and quantization.
But in recent years, there have major been strides in expanding the scope of edge systems.
Interest is brewing in both academia~\cite{sparse, hello-edge} and industry~\cite{gap8, dataset_speech} towards expanding the scope of edge ML to microcontroller-class devices.


The goal of ``TinyML'' \cite{tinymlSummit} is to bring ML inference to ultra-low-power devices, typically under a milliWatt, and thereby break the traditional power barrier preventing widely distributed machine intelligence. 
By performing inference on-device, and near-sensor, TinyML enables greater responsiveness and privacy while avoiding the energy cost associated with wireless communication, which at this scale is far higher than that of compute~\cite{warden-why-tiny}. Furthermore, the efficiency of TinyML enables a class of smart, battery-powered, always-on applications that can revolutionize the real-time collection and processing of data.
This emerging field, which is the culmination of many innovations, is poised only further to accelerate its growth in the coming years. 


To unlock the full potential of the field, hardware software co-design is required. Specifically, TinyML models must be small enough to fit within the tight constraints of MCU-class devices (e.g., a few hundred kB of memory and limited onboard compute horsepower in the order of MHz processor clock speed), thus limiting the size of the input and the number of layers~\cite{hello-edge} or necessitating the use lightweight, non-neural network-based techniques~\cite{bonsai}. TinyML tools are broadly defined as anything that enables the design, mapping, and deployment of TinyML algorithms including
aggressive quantization techniques~\cite{wang2019haq},
memory aware neural architecture searches~\cite{sparse},
frameworks~\cite{tensorflow}, 
and efficient inference libraries~\cite{cmsis-nn, pulp-nn}.
Efforts in TinyML hardware include improving inference on the next generation of general-purpose MCUs~\cite{arm-helium, gap8}, developing hardware specialized for low power inference, and creating novel architectures intended only as inference engines for specific tasks~\cite{moons2018binareye}.

The complexity and dynamicity of the field obscure the measurement of progress and make dynamism design decisions intractable.
In order to enable the continued innovation, a fair and reliable method of comparison is needed.
Since progress is often the result of increased hardware capability, a reliable TinyML hardware benchmark is required.

\begin{table*}[t]
    \caption{Survey of TinyML Use Cases, Models, and Datasets}
    \label{sample-table}
    \vskip 0.15in
    \begin{center}
    \begin{small}
    \begin{sc}
    \begin{tabular}{| M{25mm} | M{40mm}| m{25mm} | M{70mm}|}
    \toprule
    \textbf{Input Type} & \textbf{Use Cases} & \textbf{Model Types} & \textbf{Datasets} \\
    \midrule
    \\
    Audio & 
    Audio Wake Words \newline Context Recognition \newline Control Words \newline Keyword Detection & 
    DNN \newline CNN \newline RNN \newline LSTM & 
    Speech Commands ~\cite{dataset_speech} \newline Audioset ~\cite{audioset} \newline ExtraSensory ~\cite{extrasens}
    \\
    \\
    \hline
    \\
    Image & 
    Visual Wake Words \newline Object Detection \newline Image Classification \newline Gesture Recognition \newline Object Counting \newline Text Recognition & 
    DNN \newline CNN \newline SVM \newline Decision Trees \newline KNN \newline Linear &
    Visual Wake Words ~\cite{vww-dataset} \newline CIFAR10 ~\cite{cifar10} \newline MNIST ~\cite{mnist} \newline ImageNet ~\cite{imagenet} \newline DVS128 Gesture ~\cite{dvs128} \newline 
    \\
    \\
    \hline
    \\
    Physiological / Behavioral Metrics & 
    Segmentation \newline Forecasting \newline Activity Detection & 
    DNN \newline Decision Tree \newline SVM \newline Linear & 
    Physionet ~\cite{physio} 
    \newline HAR ~\cite{har} 
    \newline DSA ~\cite{dsa} \newline Opportunity ~\cite{opp} \newline UCI EMG ~\cite{emg} \\
    \\
    \hline
    \\
    Industry Telemetry &
    Sensing (light, temp, etc) \newline Anomaly Detection \newline Motor Control \newline Predictive Maintenance  
    & DNN \newline Decision Tree \newline SVM \newline Linear \newline Naive Bayes & 
    UCI Air Quality ~\cite{airquality} \newline UCI Gas ~\cite{gas} 
    \newline NASA's PCoE ~\cite{pcoe}
    \\
    \\
    
    \bottomrule
    \end{tabular}
    \end{sc}
    \end{small}
    \end{center}
    \vskip -0.1in
    \label{fig:uc, m, d}
\end{table*}

In this paper, we discuss the challenges and opportunities associated with the development of a TinyML hardware benchmark.
Our short paper is a call to action for establishing a common benchmarking for TinyML workloads on emerging TinyML hardware to foster the development of TinyML applications.
The points presented here reflect the ongoing effort of the TinyMLPerf working group that is currently comprised of over 30 organizations and 75 members. 

The rest of the paper is organized as follows. 
In Section \ref{uc, m, ds}, we discuss the application landscape of TinyML, including the existing use cases, models, and datasets.
In Section \ref{hardware}, we describe the existing TinyML hardware solutions, including outlining improvements to general-purpose MCUs and the development of novel architectures.
In Section \ref{challenges}, we discuss the inherent challenges of the field and how they complicate the development of a benchmark.
In Section \ref{related work}, we describe the existing benchmarks that relate to TinyML and identify the deficiencies that still need to be filled.
In Section \ref{benchmarks} we discuss the progress of the TinyMLPerf working group thus far and describe the four benchmarks.
In Section \ref{conclusion}, we concluded the paper and discuss future work.

\section{Tiny Use Cases, Models \& Datasets}
\label{uc, m, ds}
In this section we attempt to summarize the field of TinyML by describing a set of representative use cases (Section \ref{usecase}), their relevant datasets (Section \ref{dataset}), and the model architectures commonly applied to these specific use cases (Section \ref{model}).

\subsection{Use Cases}
\label{usecase}
Despite the general lack of maturity within the field, there are a number of well established TinyML use cases.
We categorize the application landscape of tiny ML by input type in Table \ref{fig:uc, m, d}, which in the context of TinyML systems plays a crucial role in the use case definition.

Audio wake words is already a fairly ubiquitous example of always-on ML inference. Audio wake words is generally a speech classification problem that achieves very low power inference by limiting the label space, often to two labels: ``wake word" and ``not wake word"~\cite{hello-edge}. 

Anomaly detection and predictive maintenance are commonly deployed on MCUs in factory settings where audio, motor bearing, or IMU data can be used to detect faults in products or equipment.

Other deployed TinyML applications, like activity recognition from IMU data~\cite{hassan2018robust}, rely on low feature dimensionality to fit within the tight constraints of the platforms. Some use cases have been proven viable, but have yet to reach end users because they are too new, like visual wake words~\cite{vww-dataset}.

Many traditional ML use cases can be considered futuristic TinyML tasks. As ultra-low-power inference hardware continues to improve, the threshold of viability expands. Tasks like large label space image classification or object counting are well suited for low-power always-on applications but are currently too compute and memory hungry for today's TinyML hardware. 

Furthermore, TinyML has a significant role to play in future technology. For example, many of the fundamental features of augmented reality (AR) glasses are always-on and battery-powered. Due to tight real time constraints, these devices cannot afford the latency of offloading computation to the cloud, an edge server, or even an accompanying mobile device. Thus, due to shared constraints, AR applications can benefit significantly from progress in the field of TinyML.

\subsection{Datasets}
\label{dataset}
There are a number of open-source datasets that are relevant to TinyML usecases. Table \ref{fig:uc, m, d} breaks them down by the type of data. Despite the availability of these datasets, the majority of deployed TinyML models are trained on much larger, proprietary datasets. The open-source datasets that are competitively large are not TinyML specific.  The lack of large, TinyML focused, open-source datasets slows the progress of academic research and limits the ability of a benchmark to represent real workloads accurately.

\subsection{Models}
\label{model}
Table \ref{fig:uc, m, d} lists common model types for TinyML use cases. Although neural networks (NN) are a dominant force in traditional ML, it is common to use non-NN based solutions like decision trees~\cite{bonsai}, for some TinyML use cases, due to their low compute and memory requirements.

Machine learning on MCU-class devices has only recently become feasible; therefore, the community has yet to produce models that have become widely accepted as MobileNets have become for mobile devices. This makes the task of selecting representative models challenging.
However, immaturity also brings opportunity as our decisions can help direct future progress.
Selecting a subset of the currently available models, outlining the rules for quality versus accuracy trade-offs, and prescribing a measurement methodology that can be faithfully reproduced will encourage the community to develop new models, runtimes, and hardware that progressively outperform one another.


\begin{figure}[t!]
    \centering
    \includegraphics[width=.8\linewidth]{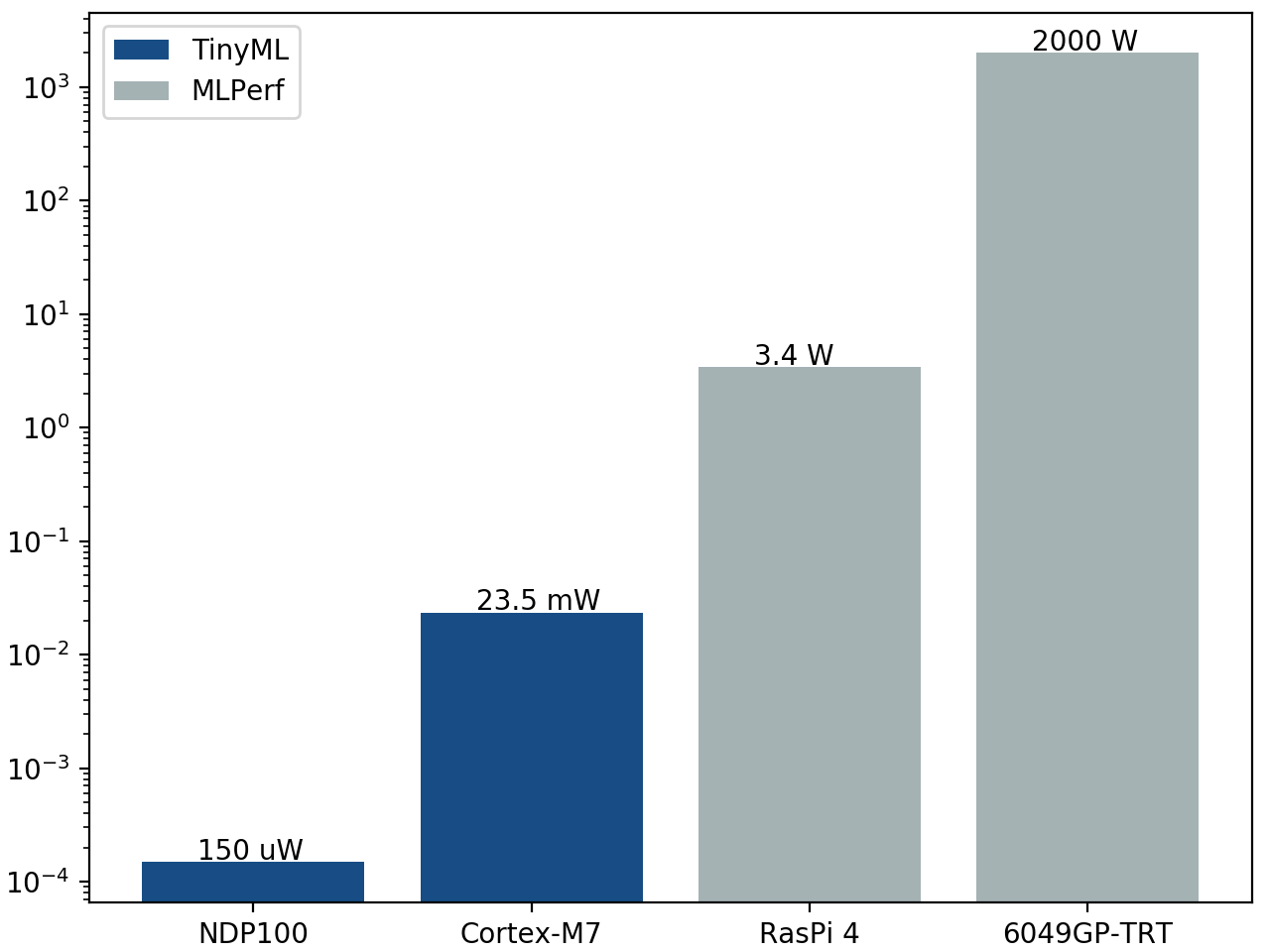}
    \caption{A logorithmic comparison of the active power consumption between TinyML systems and those supported by MLPerf.  TinyML systems can be up to four orders of magnitude smaller in the power budget as compared to state-of-the-art MLPerf systems.}
    \label{fig:tinyml vs mlperf}
\end{figure}

\section{Tiny Hardware Constraints}
\label{hardware}
TinyML hardware is defined by its ultra-low power consumption, which is often in the range of 1 mWatt and below. At the top of this range are efficient 32-bit MCUs, like those based on the Arm Cortex-M7 or RISC-V PULP processors, and at the bottom are novel ultra-low-power inference engines. Even the largest TinyML devices consume drastically less power than the smallest traditional ML devices. 
Figure \ref{fig:tinyml vs mlperf} shows the logarithmic comparison of the active power consumption between TinyML devices and those currently supported by MLPerf (v0.5 inference results from the open and closed divisions). TinyML devices can be up to four orders of magnitude smaller in the power budget as compared to state-of-the-art MLPerf systems.

The advent of low-power, cheap 32-bit MCUs have revolutionized the compute capability at the very edge. Cortex-M based platforms are now regularly performing tasks that were previously infeasible at this scale, mostly due to support for single instruction multiple data (SIMD) and digital signal processing (DSP) instructions. This fast vector math supports NN and highly efficient SVM implementations, it also accelerates many feature computations using 8bit fixed point arithmetic.

A feature of MCUs is the prevalence of on-chip SRAM and embedded Flash. Thus, when models can fit within the tight on-chip memory constraints, they are free of the costly DRAM accesses that hamper traditional ML. Widespread adoption and dispersion of TinyML are reliant on the capability of these platforms.

Although general-purpose MCUs provide flexibility, the highest TinyML performance efficiency comes from specialized hardware. Novel architectures can achieve performance in the range of one micro Joule per inference ~\cite{keyword-spotting-power-comparison}. These specialized devices expand the boundaries of ML to the ultra low power end of TinyML processors. 

\section{Challenges}
\label{challenges}
TinyML systems present a number of unique challenges to the design of a performance benchmark that can be used to measure and quantify performance differences between various systems systematically. We discuss the four primary obstacles and postulate how they might be overcome.

\subsection{Low Power}
Low power consumption is one of the defining features of TinyML systems. Therefore, a useful benchmark should ostensibly profile the energy efficiency of each device. However, there are many challenges in fairly measuring energy consumption. Firstly, as illustrated in Figure \ref{fig:tinyml vs mlperf}, TinyML devices can consume drastically different amounts of power, which makes maintaining accuracy across the range of devices difficult. 

Secondly, determining what falls under the scope of the power measurement is difficult to determine when data paths and pre-processing steps can vary significantly between devices. Other factors like chip peripherals and underlying firmware can impact the measurements. Unlike traditional high-power ML systems, TinyML systems do not have spare cores to load the System-Under-Test (SUT)  with minimal overheads.

\subsection{Limited Memory}
Due to their small size, TinyML systems often have tight memory constraints. While traditional ML systems like smartphones cope with resource constraints in the order of a few GBs, tinyML systems are typically coping with resources that are two orders of magnitude smaller. 

Memory is one of the primary motivating factors for the creation of a TinyML specific benchmark. Traditional ML benchmarks use inference models that have drastically higher peak memory requirements (in the order of gigabytes) than TinyML devices can provide. This also complicates the deployment of a benchmarking suite as any overhead can significantly impact power consumption or even make the benchmark too big to fit. Individual benchmarks must also cover a wide range of devices; therefore, multiple levels of quantization and precision should be represented in the benchmarking suite. Finally, a variety of benchmarks should be chosen such that the diversity of the field is supported.



\subsection{Hardware Heterogeneity}
Despite its nascency, TinyML systems are already diverse in their performance, power, and capabilities.
Devices range from general-purpose MCUs to novel architectures, like in event-based neural processors~\cite{brainchip-holdings-ltd} or memory compute~\cite{kim20191}.
This heterogeneity poses a number of challenges as the system under test (SUT) will not necessarily include otherwise standard features, like a system clock or debug interface. Furthermore, the task of normalizing performance results across heterogeneous implementations is a key challenge.

Today's state-of-the-art benchmarks are not designed to handle the challenges readily. They need careful re-engineering to be flexible enough to handle the extent of hardware heterogeneity that is commonplace in the TinyML ecosystem. 

\subsection{Software Heterogeneity}
There are three distinct methods for model deployment on to TinyML systems: hand coding, code generation, and ML interpreters.

Hand coding often produces the best results as it allows for low-level, application specific optimizations; however, the task is time consuming and the impact of the optimizations are often opaque to anyone but the original design team. Moreover, hand coding limits the ability to share knowledge and adopt new methods, which is detrimental to the rate of progress in TinyML. From a benchmarking perspective, hand coded submission will likely produce the best numerical results at the cost of reproducibility, comparability and time.

Code generation methods produce well optimized code without the significant effort of hand coding by abstracting and automating system level optimizations. However, code generation does not address the issues with comparability, as each major vendor has their own set of proprietary tools and compilers, which also makes portability a challenge.

ML interpreters allow for significant portability as their abstract structure is the same across platforms. TensorFlow Lite for Microcontrollers, a popular ML framework for TinyML, uses an interpreter to call individual kernels, like convolution, during run time. The framework is independent of the model architecture, therefore new models can be easily swapped in. Additionally, the reference kernels can be individually optimised and changed to fit the platform. This method comes with a small overhead in binary size and performance. From a benchmarking perspective, this abstraction separates the impact of the model architecture on the system level performance, which makes results more generalizable.

A benchmark suite must balance optimality with portability, and comparibility with representativeness. A TinyML benchmark should support many options for model deployment but the impact of that choice on the results must be carefully evaluated.

\section{Related Work}
\label{related work}
There are a number of ML related hardware benchmarks, however, none that accurately represent the performance of TinyML workloads on tiny hardware. Table~\ref{fig:benchmarks} shows a sampling of the widely accepted industry benchmarks that are directly applicable to the discussion on TinyML systems.

EEMBC CoreMark~\cite{coremark} has become the standard performance benchmark for MCU-class devices due to its ease of implementation and use of real algorithms. Yet, CoreMark does not profile full programs, nor does it accurately represent machine learning inference workloads. 

EEMBC MLMark~\cite{mlmark} addresses these issues by using actual ML inference workloads. However, the supported models are far too large for MCU-class devices and are not representative of TinyML workloads. They require far too much memory (GBs)  and have significant run times. Additionally, while CoreMark supports power measurements with ULPMark-CM~\cite{ulpmark}, MLMark does not, which is critical for a TinyML benchmark.

MLPerf, a community-driven benchmarking effort, has recently introduced a benchmarking suite for ML inference~\cite{ml_perf_inference} and has plans to add power measurements. However, much like MLMark, the current MLPerf inference benchmark precludes MCUs and other resource-constrained platforms due to a lack of small benchmarks and compatible implementations.

As Table~\ref{fig:benchmarks} summarizes, there is a clear and distinct need for a TinyML benchmark that caters to the unique needs of ML workloads, makes power a first-class citizen and prescribes a methodology that suits TinyML.

\begin{table}[t]
\caption{Existing Benchmarks}
\label{sample-table}
\begin{center}
\begin{small}
\begin{sc}
\begin{tabular}{p{4cm}ccc}
\toprule
\textbf{Benchmark} & \textbf{ML?} & \textbf{Power?} & \textbf{Tiny?} \\
\midrule
CoreMark &$\times$ & $\surd$ & $\surd$ \\
MLMark & $\surd$ & $\times$ & $\times$ \\
MLPerf Inference & $\surd$ & $\surd$ & $\times$ \\
TinyML Requirements & $\surd$ & $\surd$ & $\surd$ \\

\bottomrule
\end{tabular}
\end{sc}
\end{small}
\end{center}
\vskip -0.1in
\label{fig:benchmarks}
\end{table}

\section{Benchmarks}
\label{benchmarks}
To overcome theses challenges, we adopt a set of principles for the development of a robust TinyML benchmarking suite and select a set of 4 benchmarks.

\subsection{Open and Closed Divisions}
As previously stated, TinyML is a diverse field, therefore not all systems can be accommodated under strict rules, however, without strict rules, direct comparison of the hardware becomes more difficult. To address this issue, we adopt MLPerf's open and closed structure. More traditional TinyML solutions can submit to the closed division where submissions must use a model that is considered equivalent to the reference model.
TinyML systems that fall outside the bounds of the "closed" benchmark can submit results to the open division which will allow submissions to deviate as necessary from the closed reference. We believe this structure increases the inclusivity of the bechmarking suite while maintaining the comparability of the results.

Additionally, the open division allows for submissions to demonstrate novel software optimizations. Software based organizations can submit results using the reference platform while altering the model or inference engine to demonstrate the relative advantage of their unique solutions.

\subsection{Use Cases}
Our use case selection process prioritized diversity, feasibility, and industry relevance. Diversity to ensure our benchmark suite covered as much of the field as possible, feasibility in terms of access to open source datasets and models, and relevance to real world applications.

The group has selected four use cases to target: audio wake words,  visual wake words, image classification, and anomaly detection.
Audio wake words refers to the common, keyword spotting task (e.g. ``Alexa", ``Ok Google", and ``Hey Siri").
Visual wake words is a binary image classification task that indicates if a person is visible in the image or not.
The image classification use cases targets small label set size image classification.
Anomaly detection is a broader use case that classifies time series data as ``normal" or ``abnormal". We specifically select audio anomaly detection as our use case due to the availability of a relevant dataset.

These use cases have been selected to represent the broad range of TinyML. They encompass three distinct input data types and range from relatively resource hungry (visual wake words) to light weight (anomaly detection). Furthermore the models traditionally used for these use cases are varied therefore the benchmarking suite can support a diverse set of ML techniques.

\begin{table*}[t]
    \caption{TinyMLPerf Benchmarking Suite}
    \label{usecase_dataset_model_table}
    \vskip 0.15in
    \begin{center}
    \begin{small}
    \begin{sc}
    \begin{tabular}{ M{30mm}| M{60mm} | M{60mm}|}
    \toprule
    \textbf{Use Case} & \textbf{Datasets} & \textbf{Model} \\
    \midrule
    \\
    Audio Wake Words& 
    Speech Commands \cite{dataset_speech}&
    DS-CNN ~\cite{hello-edge}
    \\
    \\
    \hline
    \\
    Visual Wake Words &
    Visual Wake Words Dataset ~\cite{vww-dataset} &
    DS-CNN \cite{tflm-person-detection}
    \\
    \\
    \hline
    \\
    Image Classification &
    CIFAR10\cite{krizhevsky2009learning} &
    Resnet 8 \cite{he2016deep}
    \\
    \\
    \hline
    \\
    Anomaly Detection &
    ToyADMOS (Toy Car)\cite{koizumi2019toyadmos} &
    Deep AutoEncoder \cite{Koizumi_DCASE2020_01}
    \\
    \\
    
    \bottomrule
    \end{tabular}
    \end{sc}
    \end{small}
    \end{center}
    \vskip -0.1in
    \label{fig:uc, m, d}
\end{table*}

\subsection{Dataset Selection}
The group has selected a dataset for each use case, as shown in Table \ref{usecase_dataset_model_table}. The datasets help specify the use cases, are used to train the reference models, and are sampled to create the tests sets used during the measurement on device. Furthermore, the datasets can be used to train a new or modified model in the open division. We have selected datasets that are open, well known, and relevant to industry use cases.

\subsection{Model Selection}
The group has selected four reference models. These reference models are the benchmark workloads in the closed division and act as a baseline for the open division. The DS-CNN described in \cite{hello-edge} have been selected for audio wake words. 
The MobilenetV1\cite{howard2017mobilenets} used in the TensorFlow Lite for Microcontrollers person detection example \cite{tflm-person-detection} has been selected for visual wake words.
An eight layer ResNet model \cite{he2016deep} has been selected for image classification.
The baseline deep autoencoder from Task 2 of DCASE2020 competetition has been selected for anomaly detection. \cite{Koizumi_DCASE2020_01}. 
The models were selected, based on industry input, to be representative of their respective use cases. 

\subsection{Metrics}
The benchmarking suite will primarily measure inference latency with the option to measure energy consumption. The scope of the the the measurements is determined by each benchmark. In the open division the accuracy of the model must remain within a set threshold of the closed division model.

\subsection{Future work}
Perfection is often the enemy of good, therefore, to fill the community's need for comparability, our priority is to quickly establish a set of minimum viable benchmarks and iteratively address deficiencies. The benchmarking suite will continue to evolve to meet the needs of the community. 

We plan to accept result submissions in March of 2021.

\section{Conclusion}
\label{conclusion}
In conclusion, TinyML is an important and rapidly evolving field that requires comparability amongst hardware innovations to enable continued progress and stability. In this paper, we reviewed the current landscape of TinyML, including highlighting the need for a hardware benchmark. Additionally, we analyzed challenges associated with developing said benchmark and discussed a path forward. Finally, we have selected use cases, datasets, and models for our four benchmarks.

If you would like to contribute to the effort, join the working group here: \url{https://groups.google.com/u/4/a/mlcommons.org/g/tiny}

The benchmark suite is available here: \url{https://github.com/mlcommons/tiny}





\bibliography{tinyMLPerf-MLSYS}{}
\bibliographystyle{sysml2019}

%


\end{document}